\documentclass{article}
\usepackage{spconf,amsmath,amssymb,graphicx,booktabs}
\usepackage[hidelinks,pdftitle={How Much of Cross-Subject EEG Emotion Recognition Is Stimulus Memorization?},pdfauthor={Haiyan Xu and Konstantinos N. Plataniotis}]{hyperref}
\usepackage{eso-pic}
\AddToShipoutPictureBG*{\AtPageLowerLeft{\put(\LenToUnit{19mm},\LenToUnit{11mm}){\parbox[b]{178mm}{\centering\scriptsize This work has been submitted to the IEEE for possible publication. Copyright may be transferred without notice, after which this version may no longer be accessible.}}}}
\usepackage{xcolor}
\newif\ifdraftnotes
\draftnotesfalse
\ifdraftnotes
  \newcommand{\stub}[2]{%
    \par\vspace{1pt}\noindent\fcolorbox{gray}{gray!8}{%
    \parbox{0.955\columnwidth}{\scriptsize\textsc{stub (#1).}~#2}}\par\vspace{1pt}}
\else
  \newcommand{\stub}[2]{}
\fi

\makeatletter
\let\spconf@thebibliography\thebibliography
\def\thebibliography#1{\small\spconf@thebibliography{#1}%
  \setlength{\itemsep}{0pt plus 0.3pt}%
  \setlength{\parsep}{0pt}%
  \setlength{\topsep}{3pt}}
\g@addto@macro\normalsize{%
  \setlength{\abovedisplayskip}{4pt plus 1pt minus 2pt}%
  \setlength{\belowdisplayskip}{4pt plus 1pt minus 2pt}%
  \setlength{\abovedisplayshortskip}{0pt plus 1pt}%
  \setlength{\belowdisplayshortskip}{3pt plus 1pt minus 1pt}}
\makeatother

\newcommand{\Apr}{A_{\mathrm{pr}}}
\newcommand{\Dps}{\Delta_{\mathrm{ps}}}
\newcommand{\Dlop}{\Delta_{\mathrm{LOPMO}}}

\title{How Much of Cross-Subject EEG Emotion Recognition \\ Is Stimulus Memorization?}

\name{Haiyan Xu, Konstantinos N. Plataniotis}
\address{The Edward S. Rogers Sr. Department of Electrical and Computer Engineering\\
         University of Toronto, Toronto, ON M5S 3G4, Canada}

\begin{document}
\maketitle

\begin{abstract}
Cross-subject EEG emotion recognition now reports accuracies above 90\% on stimulus-defined
benchmarks such as SEED, and the field has examined several ways in which such numbers can be
inflated, attributing inflation to subject identity, to data partitioning, and to repeated stimulus
presentation. We observe that one variable has been left out. On these datasets the emotion label
is assigned to the video at design time, so any encoder that recovers which video was shown obtains
the label for free. We adapt three representation-level shortcut diagnostics to a cohort of
published contrastive cross-subject encoders, and find that stimulus identity is linearly
recoverable at six to fifteen times chance. We then measure its impact on the reported accuracy by
holding out subjects and stimuli jointly, and find that each released classifier loses two to
forty-four points once the test videos are unseen, so a headline obtained under subject hold-out
alone is not a measure of cross-subject emotion decoding.
\end{abstract}


\section{Introduction}
\label{sec:intro}

Contrastive self-supervised pretraining has become the dominant paradigm for cross-subject EEG
emotion recognition on stimulus-paradigm datasets. CLISA \cite{shen2022clisa} introduced
cross-subject same-stimulus positive pairs as an inductive bias for subject-invariant features, a
design refined by MSHCL \cite{chang2025mshcl} and carried into a dynamic-attention state-transition
encoder by DAEST \cite{shen2025daest}. Reported accuracies have risen accordingly, and the most recent
benchmarks continue the trend, with CSCL \cite{alghamdi2025cscl} and SS-EMERGE \cite{ahuja2025ssemerge}
reporting 97.70\% and 92.35\% on SEED \cite{lu2015seed,zheng2015seed}.

The field has not accepted such numbers uncritically, and has examined at least three ways in which
they can be inflated. Ahuja and Sethia \cite{ahuja2025ssemerge} warn that an evaluation protocol
``may lead to inflated performance due to subject-specific feature leakage.'' Moreno-Alcayde et
al.~\cite{morenoalcayde2024video} show that accuracy on DEAP \cite{koelstra2012deap} moves between
roughly 50\% and 90\% depending only on how the data are partitioned, and Bendrich et
al.~\cite{bendrich2022lopmo} reach a similar conclusion on AMIGOS \cite{mirandacorrea2021amigos} by comparing four cross-validation
frameworks. In an adjacent domain, Kilgallen et al.~\cite{kilgallen2024exemplar} identify a
repeated-stimulus confound in single-trial visual decoding, in which the same stimulus is presented
repeatedly to a participant and appears in both training and test.

Each of these accounts localizes inflation to one variable: subject identity, the partition, or
stimulus repetition. The most recent survey of cross-subject EEG decoding consolidates this view:
for emotion recognition, because the subject does not reveal the affective state, ``the identity
shortcut is not available'' and the remaining difficulty is domain shift \cite{li2026crosssubject}.

\begin{figure*}[t]
\centering
\includegraphics[width=\textwidth]{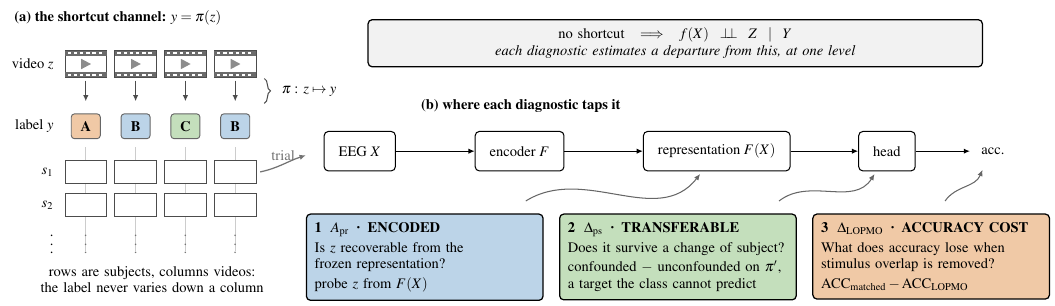}
\caption{The shortcut channel and where each diagnostic taps it. (a) One emotion is
assigned per video. (b) Each diagnostic
estimates a departure from the target \eqref{eq:ci} at a different level of the pipeline:
whether stimulus identity is encoded, whether it transfers across subjects, and what it
costs the reported accuracy.}
\label{fig:overview}
\end{figure*}

We observe that one variable is missing from this account, and that it is a property of the label
rather than of the split. On SEED and FACED \cite{chen2023faced} the
emotion class is assigned to each video when the dataset is constructed, so the label is a
deterministic function $y = \pi(z)$ of stimulus identity $z$, and an encoder that recovers which
video was shown obtains the label without representing the affective state at all. Standard cross-subject evaluation
cannot separate the two, since the held-out subject watches the same videos and those videos carry
their labels with them, so we read a reported accuracy as an unresolved mixture of
cross-subject emotion decoding and stimulus memorization. The partition accounts could not have detected this channel on the
data they used, since DEAP labels are self-reported and AMIGOS labels are rater-assigned, so in
neither case is the class a function of the stimulus.

Measuring the mixture requires diagnostics that inspect the representation and not only the split,
and although such tools are established in adjacent fields
\cite{belinkov2022probing,elazar2021amnesic,ravfogel2020inlp,kilgallen2024exemplar},
a review of 81 representation-learning studies in brain--computer interfaces reports that ``only a
few studies have introspected their learned representations'' \cite{guetschel2024review}, and we
find no prior application of them to this task.

Our contributions are twofold: (i) we characterize
the stimulus-to-label map as a shortcut channel particular to stimulus-defined datasets and establish,
across a cohort of published contrastive cross-subject encoders, that stimulus identity is present and
large in every one; and (ii) we measure its impact on the reported accuracy directly, holding out subjects
and stimuli jointly to report how much of each method's accuracy does not survive. This work
is the first systematic application of these diagnostics to cross-subject EEG emotion
recognition, and it places a measured lower bound on how much of a reported cross-subject accuracy
is stimulus memorization rather than emotion decoding.

\section{Shortcut Diagnostics}
\label{sec:diag}

\subsection{The stimulus-to-label map and the target property}
\label{sec:diag-map}
\label{sec:diag-target}

Let $z \in \{1,\dots,V\}$ index the video shown on a trial and let $y$ be the emotion label. On a
stimulus-defined dataset the two are related by a fixed map $y = \pi(z)$ chosen by the dataset
designers (Fig.~\ref{fig:overview}a), so that stimulus identity is a sufficient statistic for the label. Two structural
conditions then make the shortcut available to a contrastive encoder: positive pairs are drawn from
different subjects viewing the same stimulus, which supplies an implicit supervised signal, and the
map $\pi$ is deterministic, which makes that signal label-aligned. We regard this as the EEG-emotion
instance of shortcut learning \cite{geirhos2020shortcut}, invisible to cross-subject evaluation because holding out a
subject leaves $\pi$ intact.

Each diagnostic below estimates a departure from one target property, which is therefore stated
first. Writing $f$ for the trained predictor, Veitch et al.~\cite{veitch2021counterfactual} call $f$
\textbf{counterfactually invariant} to $Z$ when swapping the video while holding the emotion fixed
leaves the prediction unchanged, $f(X(z)) = f(X(z'))$ almost everywhere for all $z,z'$. Their Theorem 3.2
gives an observable signature for the anti-causal case that applies here,
\begin{equation}
\label{eq:ci}
f(X) \perp\!\!\!\perp Z \mid Y ,
\end{equation}
which is necessary but not sufficient for the property itself \cite{veitch2021counterfactual}. We
take \eqref{eq:ci} as the target; the three diagnostics below test it at three levels of the
pipeline (Fig.~\ref{fig:overview}b). They are complementary rather than
interchangeable: none is in the units of another, and probe accuracy is known not to track an
attribute's importance to the model~\cite{elazar2021amnesic}, so we read them together and do not
compare them with one another.

\subsection{Encoded: linear stimulus-identity probe}
\label{sec:diag-probe}

To establish that the shortcut is present at all, we first verify that stimulus identity is
recoverable from the representation, independently of whether the emotion head, the classifier
trained on the encoder output, makes use of it. A
linear probe of a frozen representation is a standard information-leakage diagnostic
\cite{belinkov2022probing}. We train a logistic-regression probe $g_w^{(k)}$ on
fold $k$'s training subjects to predict $z$ from the frozen encoder output $F(X)$, and average
held-out accuracy over the $K$ folds,
\begin{equation}
\label{eq:probe}
\Apr \;=\; \frac{1}{K}\sum_{k=1}^{K}
\mathrm{Acc}\!\left(g_w^{(k)} \circ F \,;\, \mathcal{D}_{\mathrm{test}}^{(k)}\right),
\end{equation}
with chance at $1/V$. We read $\Apr$ against that chance and against the class-only ceiling of the
next section: a value above both says that the representation carries video identity, whether or not the
head exploits it.

\subsection{Cross-subject transferable: pseudocategory bias}
\label{sec:diag-pseudo}

Because the label is a deterministic function of the stimulus, a high $\Apr$ does not by itself
establish that the representation carries stimulus information beyond the emotion class: a
representation encoding the class alone would already narrow the candidate videos and score up to
$1/K_{\mathrm{ps}}$ for $K_{\mathrm{ps}}$ videos per class, rather than $1/V$. Isolating the stimulus component requires a decoding target that the emotion class
cannot predict. We adopt the pseudocategory construction of Kilgallen et
al.~\cite{kilgallen2024exemplar}, extending it from within-subject to held-out subjects and videos. A map
$\pi' : \mathcal{V} \rightarrow \{0,\dots,K_{\mathrm{ps}}-1\}$ assigns the videos to
$K_{\mathrm{ps}}$ pseudocategories under their per-class balance constraint, each pseudocategory
containing exactly one video from each true emotion class, so that above-chance decoding of $\pi'$
cannot exploit the class alone. Each fold holds out a subject block and a video block together and evaluates on a
\emph{confounded} test set (held-out subjects, seen videos) and an \emph{unconfounded} one
(held-out subjects, held-out videos), giving
\begin{equation}
\label{eq:pseudo}
\Dps \;=\; \mathrm{ACC}_{\mathrm{conf}}(\pi') - \mathrm{ACC}_{\mathrm{unconf}}(\pi') .
\end{equation}
The unconfounded set has expectation $1/K_{\mathrm{ps}}$. A large $\Dps$ indicates that the encoder
carries stimulus-identity features beyond the class that survive a change of subject and that the head exploits while
the videos remain in distribution; a $\Dps$ near zero indicates that this content does not transfer
across subjects at the tested decoding resolution.

\subsection{Accuracy cost: joint subject-and-stimulus hold-out}
\label{sec:diag-lopmo}

The probe and the pseudocategory bias establish that the shortcut is present; this diagnostic
measures how much of the reported accuracy is stimulus memorization, as the accuracy drop under a
joint subject-and-stimulus hold-out. Since subject hold-out leaves the map $\pi$ intact
(Section~\ref{sec:diag-map}), the stimulus axis has to be held out as well, following Bendrich et
al.~\cite{bendrich2022lopmo} (leave-one-person-and-movie-out, LOPMO): each fold holds out a block of subjects and a block of stimuli jointly, every
subject block crossed with every stimulus block, each block drawing equally from the classes it
covers, and the head is retrained on the
remaining data under each method's own released configuration. We report this drop as
\begin{equation}
\label{eq:lopmo}
\Dlop \;=\; \mathrm{ACC}_{\mathrm{matched}} - \mathrm{ACC}_{\mathrm{LOPMO}} .
\end{equation}
A joint hold-out also shrinks the set of training stimuli, so the reference is not the subject-only
accuracy but a matched baseline: the head is retrained on the same reduced video set and tested on
the same held-out subjects, but on videos from that training set rather than from the held-out
block, so that the two differ only in whether the test videos were seen in training. Because the
encoder is frozen and was pretrained on all stimuli, the hold-out applies at the
classifier stage alone, so the drop is a lower bound on the accuracy attributable to stimulus
memorization rather than a point estimate of it.

\subsection{The permutation null}
\label{sec:diag-null}

The analytic null for the joint hold-out is zero. That describes the estimand, not what a released
pipeline returns: a pipeline reports some drop even when the labels carry no information, so we
measure that drop directly. Permuting the training labels over subject-by-video blocks destroys the
map of Section~\ref{sec:diag-map} together with its consistency across subjects, while leaving the
fold structure, the class balance and every other property of the protocol intact, so whatever the
apparatus reports under that permutation is what the protocol manufactures on its own. We report it
as a bias magnitude rather than as a tail probability, subtract it, taking what the protocol manufactures to be the same with and without label signal,
and print the corrected drop, written $\Dlop$ throughout; this adapts the control-task construction of Hewitt and Liang, in which a
probe is scored against its own performance on randomised labels \cite{hewitt2019probing}.

\section{Experimental Setup}
\label{sec:setup}

\begin{table*}[!t]
\centering
\caption{Shortcut diagnostics for the cohort on three datasets. ACC is each method's
reproduction at its own published protocol; the diagnostic columns are measured at the
common cohort protocol. Chance is per-diagnostic and per-dataset; under $\Dps$ it is the unconfounded
term's expectation $1/K_{\mathrm{ps}}$.}
\label{tab:diag}
\scriptsize
\setlength{\tabcolsep}{3pt}
\begin{tabular}{@{}l@{\hskip 5pt}cccc@{\hskip 8pt}cccc@{\hskip 8pt}cccc@{}}
\toprule
& \multicolumn{4}{c}{FACED-2} & \multicolumn{4}{c}{FACED-9} & \multicolumn{4}{c}{SEED-3} \\
\cmidrule(lr){2-5}\cmidrule(lr){6-9}\cmidrule(lr){10-13}
Method & ACC & $\Apr$ & $\Dps$ & $\Dlop$ & ACC & $\Apr$ & $\Dps$ & $\Dlop$ & ACC & $\Apr$ & $\Dps$ & $\Dlop$ \\
\midrule
CLISA \cite{shen2022clisa}   & $69.67{\pm}2.59$ & $38.14$ & $+18.99$ & $+17.90$ & $43.46{\pm}4.73$ & $36.35$ & $+5.97$ & $+25.99$  & $85.54{\pm}6.63$ & $40.32$ & $+38.84$ & $+12.12$ \\
MSHCL \cite{chang2025mshcl}  & $75.63{\pm}5.10$ & $26.97$ & $+23.91$ & $+5.35$ & $47.20{\pm}4.99$ & $25.46$ & $+21.63$ & $+2.17$ & ---$^{\dagger}$ & ---$^{\dagger}$ & ---$^{\dagger}$ & ---$^{\dagger}$ \\
DAEST \cite{shen2025daest}   & $79.56{\pm}4.49$ & $58.54$ & $+24.52$ & $+26.12$ & $63.55{\pm}7.35$ & $55.30$ & $+13.27$ & $+44.43$ & $84.56{\pm}6.69$ & $43.89$ & $+46.63$ & $+12.19$ \\
\midrule
\emph{chance} & --- & $4.17$ & $8.33$ & $0$ & --- & $3.57$ & $33.33$ & $0$ & --- & $6.67$ & $20.00$ & $0$ \\
\bottomrule
\end{tabular}

\vspace{2pt}
{\footnotesize
$^{\dagger}$\,MSHCL on SEED-3 is blank by design; its reproduction reaches $48.19\%$ against a
published $89.3$ \cite{chang2025mshcl}. \quad Fold counts differ per diagnostic by design: $\Apr$ uses 10 folds on FACED, $\Dps$ uses 12 on FACED-2 and 10 on FACED-9, and both use 15 on SEED-3. On FACED-9 the joint hold-out leaves three of the nine classes in each LOPMO test fold, those three being the classes whose training videos were removed; we therefore pool per-class recall across folds and macro-average over all nine (balanced accuracy) in both the joint hold-out and its matched baseline, so that $\Dlop$ compares commensurable quantities.}
\end{table*}

\noindent\textbf{Datasets and protocol.} FACED \cite{chen2023faced} contains 123 subjects, each
viewing 28 clips eliciting nine emotional states, at 250\,Hz from 32 electrodes, of which the
cohort methods retain the 30 scalp electrodes and drop the two mastoid references; we use both the
stimulus-defined 9-class label (FACED-9) and its positive-versus-negative grouping with the four
neutral clips excluded (FACED-2), following Chen et al.~\cite{chen2023faced}. SEED
\cite{lu2015seed,zheng2015seed} contains 15 subjects over three sessions viewing 15 clips eliciting
positive, neutral or negative emotions, at 200\,Hz from 62 electrodes (SEED-3). The official FACED release is
used unmodified and the DAEST-released preprocessing \cite{shen2025daest} is adopted for SEED, with
1\,second segments throughout. FACED uses the official sequential 10-fold cross-subject partition
(12 subjects per fold, 15 in the last) and SEED-3 the standard 15-fold leave-one-subject-out protocol. SEED-3's released partition
indexes session-repeated trials, so the three presentations of each clip are grouped into a single
fold, and since no released partition on either dataset supplies a joint subject-and-stimulus
hold-out, the stimulus axis is constructed as in Section~\ref{sec:diag-lopmo} on both datasets. We report means across folds, with the standard deviation across folds for ACC.

\noindent\textbf{Cohort and reproduction.} Our cohort comprises three published methods, each run as
its authors released it. CLISA \cite{shen2022clisa} is the earliest cross-subject
same-stimulus contrastive encoder; MSHCL \cite{chang2025mshcl} refines the same pairing through a
multi-scale hyperbolic embedding; and DAEST \cite{shen2025daest} keeps the same pairing and
replaces the encoder with a dynamic-attention state-transition model, so the three share one
pairing objective across three encoder architectures. Whether that pairing is what places video
identity in the representation is not tested here. Every method selects its best epoch on the
held-out fold, as released.

\section{Results}
\label{sec:results}

\begin{table}[!t]
\centering
\caption{Correcting $\Dlop$, shown on FACED-2 (balanced accuracy \%, chance 50.0). Each protocol
is measured with actual and with permuted training labels; the corrected $\Dlop$ is
$\Delta$(actual) $-$ $\Delta$(permuted), applied identically to all three datasets.}
\label{tab:permnull}
\footnotesize
\setlength{\tabcolsep}{2pt}
\begin{tabular}{@{}lrrrrrrr@{}}
\toprule
& \multicolumn{3}{c}{actual labels} & \multicolumn{3}{c}{permuted labels} & \\
\cmidrule(lr){2-4}\cmidrule(lr){5-7}
Method & matched & LOPMO & $\Delta$ & matched & LOPMO & $\Delta$ & corrected \\
\midrule
CLISA & $71.72$ & $60.21$ & $+11.51$ & $55.33$ & $61.71$ & $-6.39$ & $\mathbf{+17.90}$ \\
MSHCL & $76.42$ & $74.02$ & $+2.40$  & $55.43$ & $58.37$ & $-2.94$ & $\mathbf{+5.35}$ \\
DAEST & $81.76$ & $61.01$ & $+20.75$ & $54.73$ & $60.10$ & $-5.37$ & $\mathbf{+26.12}$ \\
\bottomrule
\end{tabular}
\end{table}

\subsection{The shortcut is present, and large}
\label{sec:results-present}

As shown in Table~\ref{tab:diag}, every encoder of the cohort
carries stimulus identity at six to fifteen times chance: probe accuracies of $26.97$
to $58.54$ on FACED-2 against a chance of $4.17$, $25.46$ to $55.30$ on FACED-9 against $3.57$, and
$40.32$ to $43.89$ on SEED-3 against $6.67$. The reference that matters, however, is the class-only ceiling $1/K_{\mathrm{ps}}$ of
Section~\ref{sec:diag-pseudo}, which is $8.33$, $33.33$ and $20.00$ on the three datasets. On FACED-2
and SEED-3 every probe clears that ceiling by two to seven times, so we read the
representation as carrying more than the class; on FACED-9, where the ceiling is highest, the probes
clear it by less than two, $36.35$ and $55.30$ against $33.33$, and MSHCL's $25.46$ not at all, so there it is the pseudocategory bias that
carries the attribution. That bias is positive on every cell, from $5.97$ to $24.52$ on FACED and $38.84$ and $46.63$ on
SEED-3, so the video-identity features the head exploits survive a change of subject
(Section~\ref{sec:diag-pseudo}). In every encoder of the cohort, then, stimulus identity is present
and large: what the classifier has is the video itself, not merely its class.

\subsection{How much of the accuracy is stimulus memorization}
\label{sec:results-lopmo}

That stimulus identity is present in the representation does not establish how much of the
reported accuracy depends on it. The $\Dlop$ column of Table~\ref{tab:diag} is an accuracy drop, the matched-baseline accuracy minus
the joint-hold-out accuracy of Section~\ref{sec:diag-lopmo} with the permutation null of
Section~\ref{sec:diag-null} subtracted, measured with each method's own classifier. The matched
baseline is the accuracy under subject hold-out alone on the reduced video set of
Section~\ref{sec:diag-lopmo}, so the drop is what such a headline loses once its test videos are unseen; it is not computed from the ACC column,
which is reproduced at each method's published protocol.

Table~\ref{tab:permnull} shows the calculation of the corrected $\Dlop$ on one dataset in full, and
its permuted-label rows show why the correction is not cosmetic: on FACED-2, with no information in
the labels, the apparatus still scores above chance on every protocol, and higher on the joint
hold-out than on the matched baseline, because the joint hold-out leaves a smaller held-out fold,
and selecting the best epoch on it gains more when the fold is small; the uncorrected estimate
therefore understates the cost. The same correction is applied on
every dataset, and since the permuted drop enters with its sign, it can cut a drop as well as add
to it.

From Table~\ref{tab:diag}, the corrected drop is positive on every cell of the cohort, from $2.17$ to
$44.43$ points, so on every dataset the released classifiers lose accuracy once the
test videos are unseen, with the subjects held out and the training set held fixed, and the drop
is therefore attributable to the videos alone. Since Section~\ref{sec:results-present} shows the features the
head exploits to be video-specific rather than class-level, we read that drop as memorization. The
drop is largest in DAEST on both FACED cells and positive under all three architectures, so the
shortcut is not particular to one encoder architecture. Part of each method's accuracy, up to
forty-four points, therefore does not survive a change of videos, and a headline obtained under subject hold-out alone on a stimulus-defined dataset is
not a measure of cross-subject emotion decoding.

\section{Conclusion}
\label{sec:conclusion}

In this paper, we investigated how much of a reported cross-subject EEG emotion accuracy is stimulus
memorization when the label is fixed by the video, by adapting three representation-level shortcut
diagnostics to three published contrastive encoders on FACED and SEED. Every encoder
carries stimulus identity at six to fifteen times chance, and what it carries is the individual video
rather than its emotion class. Shortcut accounts of this task have concerned the subject, the partition
and stimulus repetition; we show that stimulus memorization is a further shortcut, one that the partition
accounts could not have seen on their data. Under a joint subject-and-stimulus hold-out, each
released classifier loses between two and forty-four points of accuracy once the permutation null is
subtracted, a lower bound since the frozen encoders saw every video. A headline obtained under subject
hold-out alone on a stimulus-defined dataset therefore overstates cross-subject emotion decoding, and the three
diagnostics report, beside any such headline, how much of it survives. Future cross-subject results
should be reported on both stimulus-defined datasets and datasets without a stimulus-to-label map, such
as DEAP and AMIGOS, on which the label is not fixed by the video.

\section{Compliance with Ethical Standards}
\label{sec:ethics}

This research study was conducted retrospectively using human subject data made available in
open access by the FACED \cite{chen2023faced} and SEED \cite{lu2015seed,zheng2015seed} dataset
providers. Ethical approval was not required as confirmed by the licenses attached with the open
access data. No new human-subject data were collected for this study.

\section{Acknowledgments}
\label{sec:ack}

Work is partially supported by an NSERC Alliance Grant entitled: Objective Measurement of Trust in
Human-AI Teaming Towards Trust-Adaptive Intelligent Systems.

\bibliographystyle{IEEEbib}
\bibliography{strings,references}

\end{document}